\documentclass[12pt, preprint]{aastex}

\begin{document}

\title{\bf {\it RHESSI} Observations of a Simple Large X-ray Flare on 11-03-2003}

\author{Wei Liu\altaffilmark{1}, Yan Wei Jiang\altaffilmark{1}, Siming Liu\altaffilmark{1}, 
and Vah\'{e} Petrosian\altaffilmark{2}}

\affil{Center for Space Science and Astrophysics, Stanford University, Stanford, California 94305}

\altaffiltext{1}{Department of Physics; weiliu@sun.stanford.edu, arjiang@stanford.edu, 
liusm@stanford.edu}
\altaffiltext{2}{Departments of Physics and Applied Physics; vahe@astronomy.stanford.edu}

\begin{abstract}

We present data analysis and interpretation of a simple X-class flare observed with {\it
RHESSI} on November 3, 2003.  In contrast to other X-class flares observed previously, this 
flare shows a very simple morphology with well defined looptop (LT) and footpoint (FP) 
sources.  The almost monotonic upward motion of the LT source and increase in separation of 
the two FP sources are consistent with magnetic reconnection models proposed for 
solar flares.  In addition, we find that the source motions are relatively slower during the 
more active phases of hard X-ray emission; the emission centroid of the LT source shifts 
toward higher altitudes with the increase of energy; the separation between the LT emission 
centroids at two different photon energies is anti-correlated with the FP flux. Non-uniformity 
of the reconnecting magnetic fields could be a possible explanation of these features.

\end{abstract}


\section{Introduction}

With its high temporal, spatial, spectral resolution and broad energy coverage, the {\it
Reuven Ramaty High Energy Solar Spectroscopic Imager} ({\it RHESSI}) has revealed many 
features of solar flares with unprecedented details (Lin et al. 2002). Since its launch on 5 
February 2002, {\it RHESSI} has observed several X-class flares and thousands of mid-class 
and small flares.  The compactness of microflares limits our access to details of the
energy release and particle acceleration processes (Krucker et al. 2002). 
On the other hand large and well resolved flares usually involve multiple loops with 
complex structures and the looptop (LT) and associated footpoint (FP) sources are not 
readily identified and separated (Gallagher et al. 2002; Lin et al.  2003).  This makes a 
direct comparison of theoretical models with observations a challenging task 
(Alexander \& Metcalf 2002; Sui et al. 2002). This task would be easier for a large flare with a 
simple morphology, where one can identify source positions and evolutions with certainty 
(Tsuneta et al. 1992; Tsuneta 1996; Tsuneta et al. 1997).

In late October and early November 2003, {\it RHESSI} and other instruments observed a series of
X-class flares from solar active regions AR 0486 and 0488 (reminiscence of the June 1991 flares
of the previous solar cycle; Schmieder et al. 1994). Among these flares, we studied an event
which occurred on November 3 in AR 0488 at heliographic coordinate N09$^\circ$ W77$^\circ$.
Unlike other X-class flares, e.g. the April 21, 2002 flare (Gallagher et al. 2002)  and the
gamma-ray flare on July 23, 2002 (Lin et al. 2003), this flare shows a surprisingly simple
morphology with well defined LT and two FP sources.

In this letter we present a brief description of the spatial evolution of the various emission 
regions of this flare.  As we will show this provides an excellent example
of the classical solar flare model of magnetic reconnection and energy release in an
inverted Y magnetic field configuration (Kopp \& Pneuman 1976; Forbes \& Acton 1996;
Aschwanden 2002), whereby reconnection in the oppositely directed field lines leads to
particle acceleration near the LT.  The energy release and particle 
acceleration processes are not well understood, nevertheless, it is expected that the 
reconnection will produce closed loops at lower altitudes first and progress to higher overlying 
loops as time advances.  Consequently, the altitude of the LT source and the separation of the 
two FPs should increase with time.  The flare studied here shows this exact behavior. 

On the other hand, we also see evidence for deviations from the simplest reconnection models. 
Our study indicate that the reconnecting fields could be nonuniform and may have a shearing 
component. In the next section, we present the observations, data analysis, and our results.  
Their implications are discussed in \S\ \ref{dis}.

\section{Observations and Data Analysis}
\label{obs}

The flare under study, classified as a {\it GOES} X3.9 class flare, was observed by {\it 
RHESSI}, {\it Solar and Heliospheric Observatory} ({\it SOHO}), etc.  Figure 1a 
shows the {\it RHESSI} light curves.  In lower energy channels ($<25$ keV), the count rates 
started to rise at around 09:43 UT, peaked about nine minutes later, and then began a monotonic 
declining phase till 10:01:20 UT when {\it RHESSI} entered the Earth's night region. 
The higher energy channel ($>50$ keV) light curves exhibit two broad impulsive bursts, each of 
them consisting of several pulses, with a more quiescent part in between, suggesting a 
persistent but episodic energy release process. Impulsive radio activities were also observed by 
the Nan\c cay Observatory (Vilmer, private communication). A partial halo Coronal Mass Ejection 
(CME) with a speed of $\sim 1375$ km/s was observed by the Large Angle and Spectrometric 
Coronagraph on {\it SOHO}.

To study the hard X-ray (HXR) source motion and structure, we obtained images at 
different energies in 20-second intervals from 09:46:20 through 10:01:00 UT using the CLEAN 
algorithm (Hurford et al. 2002) and front segments of detectors 3-8 to achieve a FWHM of 
$9\farcs8$ with a $0\farcs5$ pixel size. Figure \ref{motions.ps} shows the HXR emission contours 
during the two main activity peaks. There are three sources: a LT, a Northern FP (N-FP), and a 
Southern FP (S-FP).  The LT source dominates at lower energy while the FPs dominate the higher 
energy emission. As evident from the background pre-flare magnetogram obtained with the 
Michelson Doppler Imager (MDI), the N-FP is around a negative magnetic polarity region while 
the S-FP remains in a region of positive polarity.  Note that early in the 
event there is a partial overlap between the N-FP and the LT source. Grids with higher spatial 
resolution will not help for this flare because grid 2 has severely degraded conditions (Smith et 
al. 2002) and grid 1 will over-resolve the sources (See Schmahl \& Hurford 
2003 for technical details). A post-flare (10:35:43 UT) Extreme ultraviolet Imaging Telescope 
(EIT) 195 {\AA} image (not shown) shows a loop structure which agrees well with the {\it RHESSI} 
sources. 

As shown in Figure \ref{motions.ps} the LT and FPs have well defined and correlated motions,
with the symbols indicating their emission centroids at different times.  The yellow dashed
line represents the main direction of the LT motion, which is roughly at a right angle to the
solar limb.  We refer to the motion along this direction as changes in altitude.  The motion
perpendicular to this direction might be due to asymmetry of the reconnecting loops or LT
motion along an arcade.  Before the rise of impulsive HXR emission, there is an apparent
downward LT motion.  This downward motion could indicate a shrinkage of newly formed loops. It
may also be due to formation of nearby sources (Krucker, Hurford, \& Lin 2003), or due to
projection effects should the LT source moves eastward along arcades of loops (Sato 2001).
Qualitatively similar features have been seen in several other flares (Krucker et al. 2003;
Sui \& Holman 2003), suggesting that this may be a common characteristic of solar flares.  
However, for the remainder of the flare duration the LT source rises systematically.  The
apparent separation of the FP sources, whenever detectable, also increases with comparable
speed. As emphasized above, this is expected in a simple continuous reconnection process which
moves up to the corona, accelerating particles and energizing plasma higher up into overlying
larger loops.

To analyze the FP motion quantitatively, one needs to take into account projection effects
because any motion and its associated uncertainty in the east-west direction are amplified by
a factor of about $\csc77^{\circ} \simeq 4.4$.  Motions in this direction are highly uncertain
and the motion of both FPs appears to have an east-west component.
Magnetic reconnection, on the other hand, is
characterized by the change in the size of newly formed loops rather than their absolute
motions. Thus one may concentrate on the relative motion of the two conjugate FPs.  In the
insert panel of Figure \ref{motions.ps}, we illustrate this relative motion by fixing the S-FP
at the origin of the coordinates and showing the relative locations of the N-FP.  The relative
motion is obviously systematic. The fact that the line tracing the location of the N-FP is not
exactly aligned with the lines connecting the two FPs shows that there is another component of
the relative motion introducing a small rotation of the plane containing the newly formed
loop.  Because this line is nearly parallel to the longitudinal line, one can ignore the
projection affects.  We will quantify the relative motion along this line and the standard
deviation of the displacement (apparently) perpendicular to this line will be used as an upper
limit for the uncertainties of this relative motion.

Figure 1b shows this relative motion of the FPs (in 50-71 keV) along with the location of the
emission centroids of the LT source in three energy bands projected onto its main direction of
motion.  As evident, the two motions are correlated and the two set of data points are nearly
parallel to each other indicating comparable velocities. To further investigate these motions we
divide the observed flare duration into four phases: a pre-impulsive phase (before 09:48:10 UT)  
when there is no significant high energy HXR emission, a rising phase (from 09:48:10 to 09:49:50
UT), a declining phase (from 09:49:50 to 09:56:50 UT), and a second active phase (from 09:56:50
till 10:01:00 UT). We then fit straight lines to each segment and determine the corresponding
average velocities.  The results are summarized in Table 1. Surprisingly, the LT velocity is
highest in the declining phase, when the X-ray emission is relatively weaker (Fig. 1c). In the 
simplest model of reconnection of {\it uniform} and oppositely directed magnetic fields, one would 
expect the opposite correlation, i.e. a higher rate of energy release when the velocity is larger.
However, this would be true if the observed HXR flux were actually proportional to the total energy 
release and if reconnection were indeed occurring in a uniform background plasma, neither one of 
which is exactly true.

Another interesting morphological evolution is the change of the centroid of the LT source
with energy. In Figure \ref{structure.ps} we show the {\it RHESSI} $75\%$ contours and
centroids at several energies superposed on an MDI continuum image showing sunspots. Compared
with the two FPs, the LT source shows a clear and systematic displacement of the centroid of
the higher energy emissions toward higher altitudes, as seen in two other flares (Sui \& 
Holman 2003;  Gallagher et al. 2002). To investigate what this separation of the LT centroids 
is related to, we looked for its correlations with other characteristics. We found an
anti-correlation between the centroid separation and the high energy (100-300 keV) count rate,
which comes mainly from the FPs (Fig. 1d). The continuous curve in Figure
\ref{correlation.ps} shows their cross-correlation function, which gives a peak correlation
coefficient of $-0.51 \pm 0.08$ with a time lag of $\Delta t=-22 \pm 39$ s. The data points (LT
separation vs HXR count rate) used for evaluating the correlation and a straight-line fit are
also shown in the same figure.

\section{Discussion}
\label{dis}

We have investigated the November 3, 2003 X3.9 flare, having a simple morphology with well 
defined LT and FP sources.  The high flux combined with the simple loop structure allows us to 
determine the spatial evolution of the LT and FP sources clearly and to compare with the simple 
reconnection models. Similar studies of flares have been limited to the investigation of the 
motion of the FPs alone (Sakao, Kosugi \& Masuda 1998; Qiu et al. 2002; Fletcher \& Hudson 2002) 
or have dealt with complex loop structures (Krucker et al. 2003; Qiu, Lee \& Gary 2003). This 
has made the comparison with models more difficult. Our analysis of {\it RHESSI} data has 
yielded several new and interesting results. 

\begin{enumerate}

\item We observe a systematic rise of the LT source and a comparable increase in the separation of 
the FPs as the flare proceeds.  This agrees very well with the canonical solar flare model of 
magnetic reconnection in an inverted Y configuration.  Similar behaviors have been reported 
previously using soft X-ray or EUV observations (\v{S}vestka et al. 1987; Tsuneta et al. 1992; 
Gallagher et al. 2002) during later thermal gradual phases of flares.  However, these 
emissions are not directly related to the impulsive particle acceleration processes (Forbes \& 
Acton 1996).

\item The LT source seems to move more slowly during the HXR peaks than during the declining 
and more quiescent phases, in apparent disagreement with reconnection of {\it uniform} and 
oppositely directed field lines, where one would expect a correlation between the velocity of 
the LT source and the energy release rate.  However, we note that the HXR flux is not a good 
proxy for the energy release rate and the magnetic fields in the reconnection region are 
likely to be nonuniform.  Stronger magnetic fields would require smaller volume of 
reconnecting fields and possibly slower motion. However, in an inhomogeneous case other 
factors like the geometry and Alfv\'{e}n velocity variation can also come into play.  
This problem needs further exploration.

\item The centroid of the LT source appears to be at higher altitudes for higher photon 
energies.  This suggests that the energy releasing process happens above the LT and that 
harder spectra, implying more efficient acceleration, are produced at higher altitudes. In the 
stochastic acceleration model by turbulence where the acceleration efficiency depends on the 
intensity of turbulence, this would indicate a decrease of the intensity with decreasing 
altitudes, presumably due to decay of turbulence away from its source at a higher altitude.

\item The above shift of the centroids decreases with the increase of HXR flux from the FPs.  
Such an anti-correlation will be difficult to produce in simple models. In the above mentioned 
model, this would imply a more homogeneous distribution of turbulence during more active 
phases.



\end{enumerate}


{\bf Acknowledgments} 

The work is supported by NASA grants NAG5-12111, NAG5 11918-1, and NSF grant ATM-0312344.
We are indebted to S. Krucker for numerous help and for valuable comments as a referee.
We also would like to thank T. Metcalf, G. Hurford for helpful discussions and 
suggestions, and are grateful to K. Tolbert, R. Schwartz, and J. McTiernan for their 
help with analyzing {\it RHESSI} data.

{}

\clearpage

\vspace {0.1in}
\begin{table}[ht]
\caption{LT velocities and FP separation speed.}
\vspace {0.05in}
\begin{tabular}{lrrrrr}
\tableline\tableline
Time range  & \multicolumn{4}{c} {LT velocities (km/s)} & FP speed(km/s) \\
             \cline{2-5} 

 (UT) & 9-12 keV & 12-15 keV  & 15-19 keV & 19-24 keV & (50-71 keV) \\
\tableline
09:46:20-09:48:10 &$ -18.3\pm 3.7 $&$ -22.5\pm 4.6 $&$ -32.5\pm 4.1 $&$ -30.8\pm 4.7 $& --- --- --- \\
09:48:10-09:49:50 &$ 3.5  \pm 3.3 $&$ 4.0  \pm 3.0 $&$ 4.7  \pm 2.6 $&$ 4.3  \pm 2.7 $&$ 29.1 \pm 11.6 $ \\
09:49:50-09:56:50 &$ 14.6 \pm 0.4 $&$ 16.5 \pm 0.2 $&$ 18.0 \pm 0.2 $&$ 20.9 \pm 0.1 $&$ 22.4 \pm 2.5  $ \\
09:56:50-10:01:00 &$ 9.3  \pm 0.9 $&$ 8.6  \pm 0.7 $&$ 6.6  \pm 0.7 $&$ 5.9  \pm 0.5 $&$ 10.4 \pm 3.6  $ \\

\tableline
\end{tabular}
\end{table}

\clearpage

\clearpage

\begin{figure}[thb]  
\epsscale{0.8}
\centerline{
\plotone{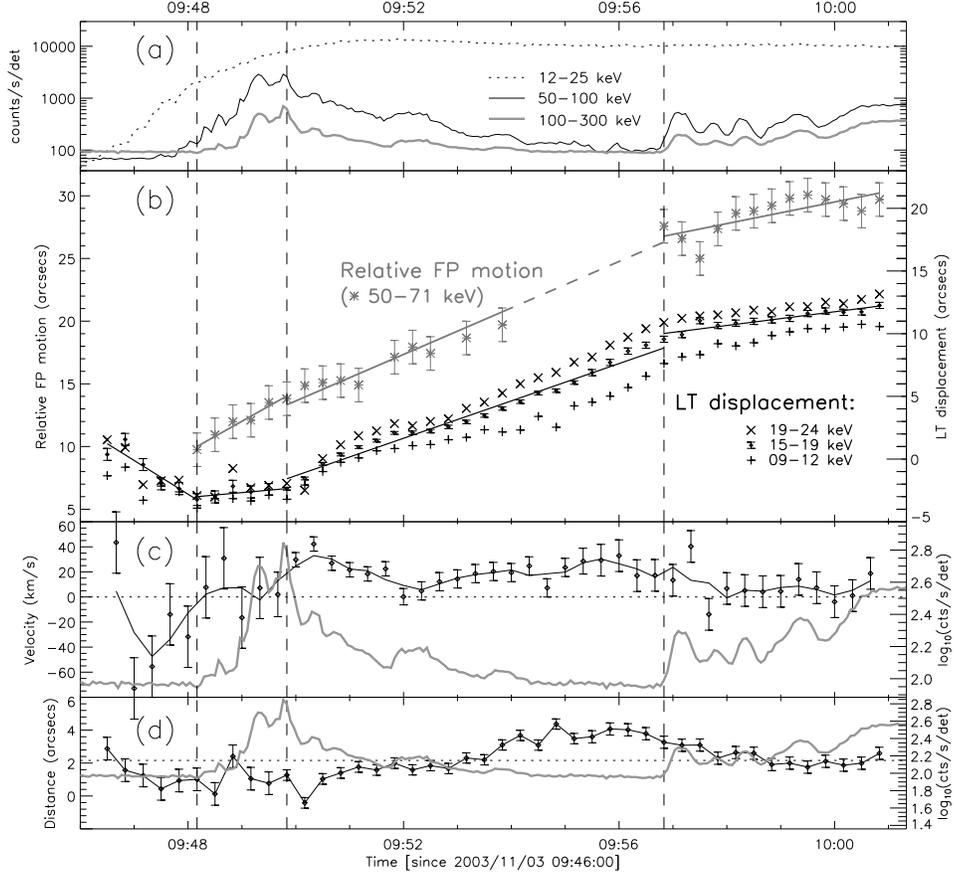}}
\caption{\small (a) {\it RHESSI} light curves (counts/second/detector). 
(b) The evolution of the altitude of the LT centroid (right scale) and the separation of 
the two FPs (left scale). The LT altitude here refers to the displacement along the main 
direction of motion which is nearly perpendicular to the solar limb. 
The straight lines are fits to the FP separation and LT altitude in 15-19 keV with vertical dashed lines 
separating the four phases as described in the text. 
The uncertainty of the centroid location in 15-19 keV is shown with the vertical error bars,
which are similar to those in 9-12 keV and 19-24 keV (not shown). The uncertainty of the 
relative FP motion is also indicated.
(c) The corresponding LT velocity in 15-19 keV. The thin curve is the velocity smoothed over 
1-minute intervals. The thick grey curve is the logarithm of the 100-300 keV count rate 
(right scale). 
(d) Separation of the LT centroids in 19-24 keV and 9-12 keV (panel b) as a function of time. 
The dotted horizontal line marks the mean of this separation and the logarithm of the count rate 
(same as panel c). 
}
\label{curves.ps}
\end{figure}

\begin{figure}[thb] 
\epsscale{0.72}
\centerline{
\plotone{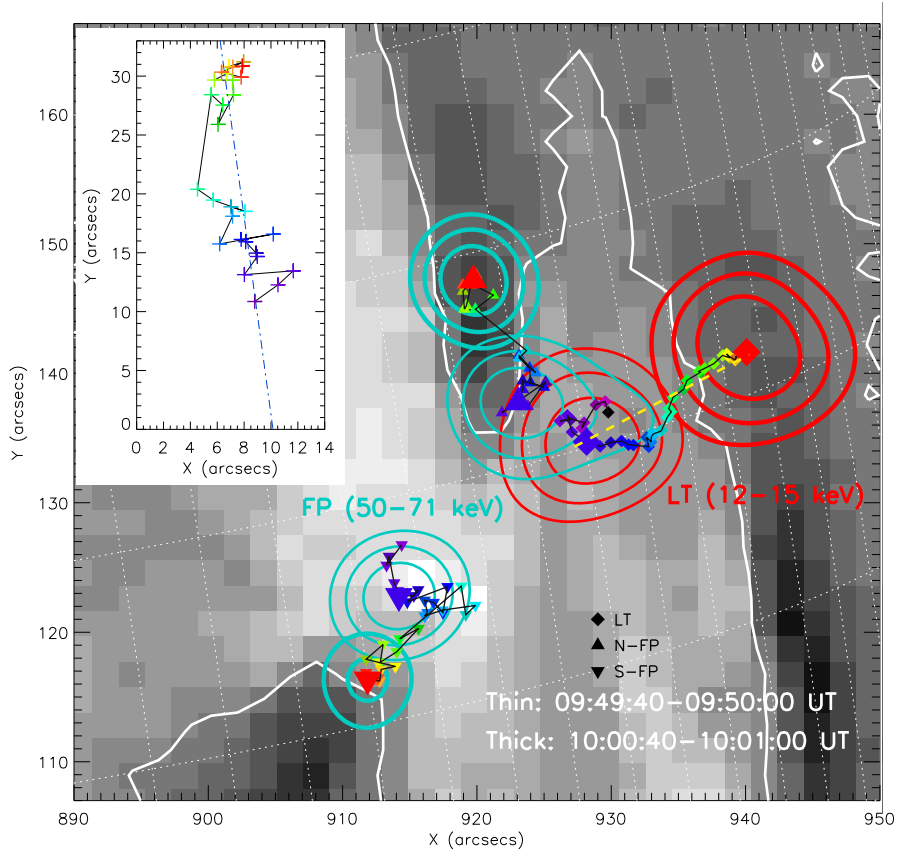}}
\caption{\small Temporal evolution of HXR source centroids, over-plotted on an MDI magnetogram 
(09:32:30 UT).  Black line segments connect the centroids obtained from CLEAN images in 
successive 20-s intervals chronologically from black 
(09:46:20 UT) through violet, blue, green, yellow, and to red (10:01:00 UT).
The LT (12-15 keV) centroid is the brightness-weighted source center within the 
70\% level contour but each FP (50-71 keV) centroid is the peak position obtained with a 
$3\times3$-pixel parabolic fit around the brightest pixel. 
The yellow dashed line represents the main direction of motion of the LT source. 
To estimate the uncertainty in the LT centroid location,
we fitted the LT data points with 4 straight lines within the 
time intervals, 09:46:20-09:49:40 UT, 09:49:40-09:52:00 UT, 09:52:00-09:55:20 UT,
and 09:55:20-10:01:00 UT, respectively. For each interval, following Krucker et al. (2003), 
the standard deviation of the offset of the data from the corresponding
straight line was used as the error in the location. The insert shows the relative positions of 
the N-FP with respect to the S-FP which is fixed at the origin. 
We attribute the motion perpendicular to the straight line to uncertainties in the 
locations (see text for details).
Four HXR images in two time intervals, 09:49:40-09:50:00 UT (thin) and 
10:00:40-10:01:00 UT (thick), and in two energy channels, 12-15 keV (red) and 50-71 keV (cyan),
are over-plotted as contours (at 55, 70, 85\% levels of the maximum brightness of the image),
which clearly depict the LT and FPs, respectively.
The centroids corresponding to these two intervals are indicated with larger symbols.
The magnetogram shows the line-of-sight magnetic field in a grey scale ranging from $-979$ 
(black: pointing away from the observer) to $+1004$ (white) Gauss.
The apparent neutral lines are marked in white. 
}
\label{motions.ps}
\end{figure}

\begin{figure}[thb]
\epsscale{1.}
\centerline{
\plotone{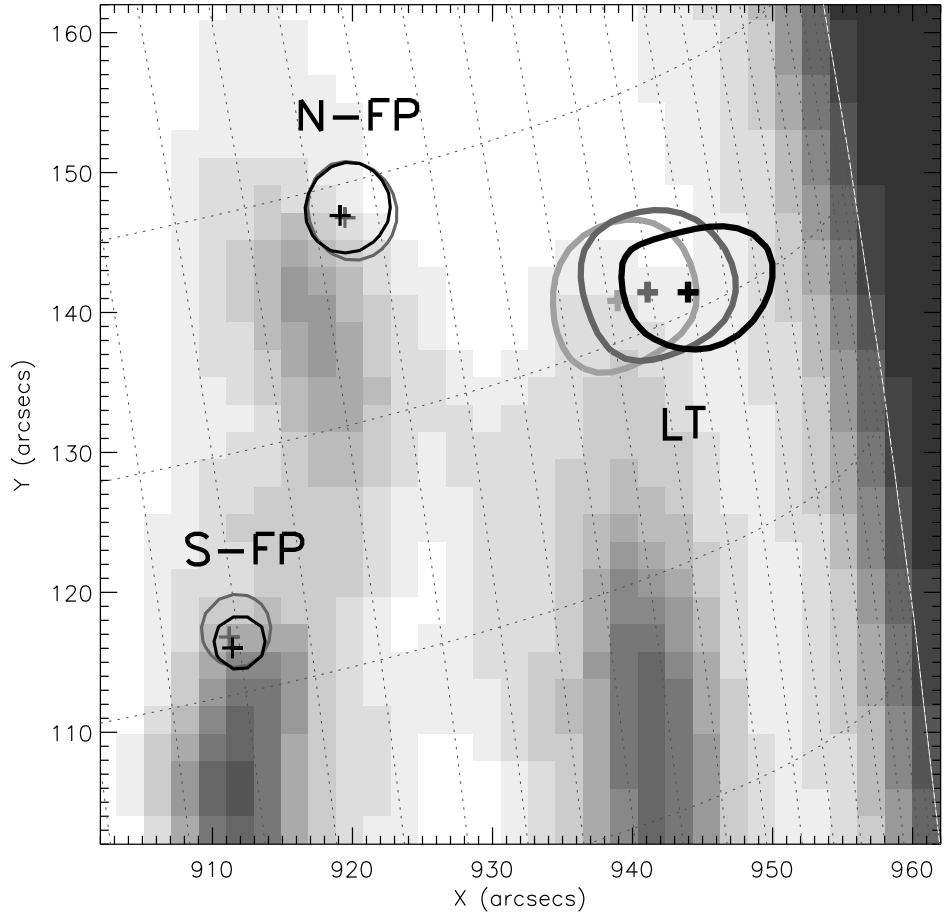}}
\caption{{\it RHESSI} image contours ($75\%$) and the corresponding brightness-weighted 
centroids (+) in the interval 10:01:00-10:01:20 UT. The LT contours are for 12-14 keV (light 
grey), 18-21 keV (grey), and 27-31 keV (dark) and the FP contours are for 40-46 keV 
(grey) and 60-73 keV (dark). The background is an MDI continuum map taken at 09:36:00 UT. 
The dark areas inside the limb are three sunspots.
}
\label{structure.ps}
\end{figure}

\begin{figure}[thb]
\epsscale{.86}
\centerline{
\plotone{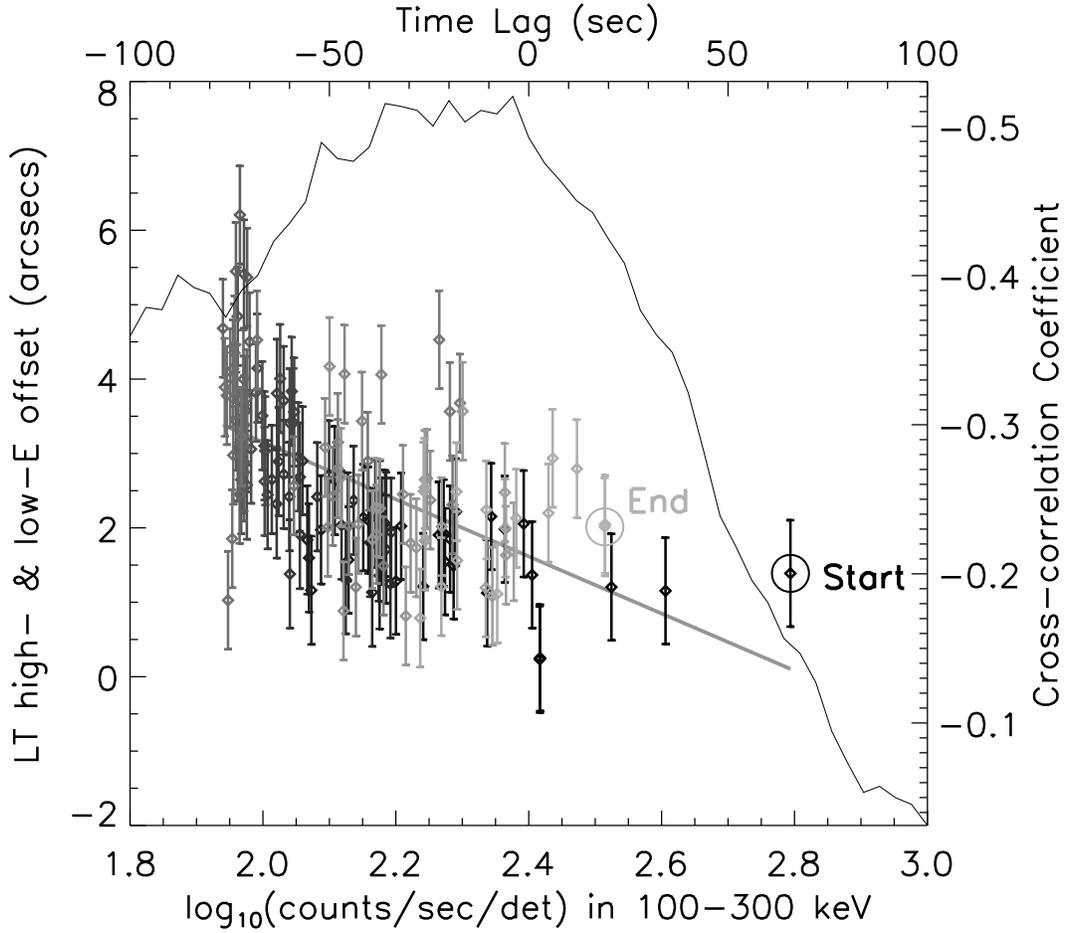}}
\caption{
Correlation between the LT structure and the 100-300 keV (mainly FPs) light curve.  
The thin curve (with the top and right axis) shows the cross-correlation coefficient 
of the logarithm of the count rate and the separation between the 19-24 keV and 9-12 keV 
centroids of the LT source, showing a $22 \pm 39$ s delay relative to the light curve. The 
separation is similar to that shown in Figure 1d but with a higher time resolution, obtained by 
imaging at a 4-s cadence (same as the light curve) with an integration time of one spacecraft 
spin period ($\sim 4$ s) from 09:49:48 to 10:01:00 UT. We excluded the first two phases of the 
flare duration when the spatial contamination to the LT source by the N-FP is severe.
The diamond symbols (with the bottom and left axis) show the LT separation versus the logarithm 
of the count rate shifted by $+24$ seconds, corresponding to the peak of 
the correlation coefficient. The vertical error bars represent the uncertainty in the centroid
separation. The darkness of the symbols represents time 
with the start and end point being circled. The 
grey thick line is a linear fit to the data with a slope of $-3.84 \pm 0.34$. 
}
\label{correlation.ps}
\end{figure}

\end{document}